\documentclass[submission, Proceedings]{SciPost}

\hypersetup{
    colorlinks,
    linkcolor={red!50!black},
    citecolor={blue!50!black},
    urlcolor={blue!80!black}
}
\usepackage[square,sort,comma,numbers]{natbib}

\usepackage{afterpage}
\usepackage{subcaption}
\usepackage[bitstream-charter]{mathdesign}
\let\OLDthebibliography\thebibliography
\renewcommand\thebibliography[1]{
  \OLDthebibliography{#1}
  \setlength{\parskip}{0pt}
  \setlength{\itemsep}{0pt plus 0.3ex}
}
\DeclareSymbolFont{usualmathcal}{OMS}{cmsy}{m}{n}
\DeclareSymbolFontAlphabet{\mathcal}{usualmathcal}

\begin{document}

\begin{center}{\Large \textbf{
PB TMD fits at NLO with dynamical resolution scale \\
}}\end{center}

\begin{center}
S. Sadeghi Barzan$i^{1,2}$
\end{center}

\begin{center}
{${ }^{1}$Department of Physics, Shahid Beheshti University, Iran}
\\
{${ }^{2}$ Elementary Particle Physics, University of Antwerp, Belgium}
\\

\end{center}


\definecolor{palegray}{gray}{0.95}
\begin{center}
  \begin{tabular}{rr}
  &
  \begin{minipage}{1.0\textwidth}
    \begin{center}
    {\it Presented at DIS2022: XXIX International Workshop on Deep-Inelastic Scattering and Related Subjects, Santiago de Compostela, Spain, May 2-6 2022} \\
    \end{center}
  \end{minipage}
\end{tabular}
\end{center}

\section*{Abstract}
{\bf
Parton branching solutions of QCD evolution equations have recently been studied
to construct both collinear and transverse momentum dependent (TMD) parton
distributions. In this formalism a soft-gluon resolution scale is introduced to 
separate resolvable and non-resolvable branchings,  and to take into account 
soft-gluon coherence effects. In 
this talk, results of fits to the high precision
deep inelastic scattering (DIS) structure function measurements are shown including for the
first time the effects of dynamical, i.e. branching-scale dependent, resolution scales at
Next-to-Leading-Order (NLO) accuracy in the strong coupling.
}


\section{Introduction}
\label{sec:intro}
QCD resummations \cite{:Luisoni} to all orders in the strong coupling are 
an essential aspect of theoretical predictions for precision physics at high-energy hadron colliders. Transverse momentum dependent (TMD) parton distributions  \cite{:R. Angeles-Martinez} provide a theoretical framework to both accomplish resummed perturbative calculations and include non-perturbative dynamics.
In Refs. \cite{Hautmann:1708, Hautmann:1704} a method which is based on the unitarity picture of parton evolution \cite{B.R. Webber, R.K. Ellis} has been presented to define TMDs in a parton branching (PB) formalism. In this method color coherence
of soft-gluon radiation \cite{A. Bassetto, Y.L. Dokshitzer, G. Marchesini, S. Catani} and transverse momentum recoils are taken into account. The soft-gluon resolution
scale is introduced to separate resolvable branchings from non-resolvable ones, and 
Sudakov form factors are used to describe explicit  partonic 
probabilities for no resolvable branchings in a given evolution interval.
Since the transverse momentum generated radiatively
in the branching is sensitive to the treatment of the non-resolvable region \cite{ F. Hautmann:hep -ph /0702196}, an additional condition can be applied to relate the transverse momentum recoil and the scale of the branching. This relation
incorporates the property of angular ordering, and implies that 
the soft-gluon resolution scale
can be dynamical, i.e., dependent on the branching scale.

In this work, the effects of dynamical resolution scales on TMD evolution and on
collider observables are discussed. The PB evolution equations are solved numerically with dynamical 
resolution scale by applying the Monte Carlo solution techniques \cite{Hautmann:1704, F. Hautmann:1407}, and fits to precision 
deep inelastic scattering (DIS) measurements \cite{Herapdf} are performed 
at Next-to-Leading-Order (NLO) accuracy in the strong coupling. 

 The paper is organized as follows. In Sec.2  the PB evolution equation and   dynamical 
 soft-gluon resolution scale are described. The results of DIS fits  with dynamical resolution scale are shown in Sec.3.  
   Conclusions are given in Sec.4.

\section{Parton Branching TMDs and angular ordering} 
In the Parton Branching approach, the TMD evolution equations can be written as \cite{Hautmann:1708} 
\begin{equation}
\begin{aligned}
\label{eqn:PB}
 \tilde A_{a}\left(x, k, \mu^{2}\right) &= \tilde A_{a}\left(x, k, \mu_{0}^{2}\right) \Delta_{a}\left(\mu^{2}, \mu_{0}^{2}\right) \\
&+\sum_{b} \int \frac{\mathrm{d}^{2}\mu'}{\pi \mu'^{2}} \Theta\left(\mu^{2}-\mu'^{2}\right) \Theta\left(\mu'^{2}-\mu_{0}^{2}\right) \\
& \times \int_{x}^{1} {dz} \Theta\left(z_{m}(\mu')-z\right)\frac{\Delta_{a}\left(\mu^{2}, \mu_{0}^{2}\right)}{\Delta_{a}\left(\mu'^{2},\mu_{0}^{2}\right)} P_{a b}^{R}\left(z, \alpha_{s}\left(b\left(z\right)^{2}\mu'^{2}\right)\right)\tilde A_{b}\left(x / z, k + a(z)\mu', \mu'^{2}\right)
\end{aligned}
\end{equation}
where $ \tilde A_{a}\left(x, k, \mu^{2}\right)= xA_{a}\left(x, k, \mu^{2}\right)$  is the momentum-weighted TMD distribution of flavor $\it{a}$, carrying the longitudinal momentum fraction $\it{x}$ of the hadron's momentum and transverse momentum $\it{k}$ at the evolution scale $\mu$; $\it{z}$ and $\mu'$ are the branching variables, with $\it{z}$ being the longitudinal momentum
transfer at the branching, and $\mu'$ the momentum scale at which the branching occurs; $P^{R}_{ab}$  are the real-emission splitting kernels; $\Delta_{a}$ is the Sudakov form factor, given by
\begin{eqnarray}
\Delta_{a}\left(\mu^{2}, \mu_{0}^{2}\right)=\exp \left(-\sum_{b} \int_{\mu_{0}^{2}}^{\mu^{2}} \frac{\mathrm{~d} \mu^{\prime 2}} {\mu'^{2}} \int_{0}^{1} \mathrm{~d} z \Theta\left(z_{m}\left(\mu'\right)-z\right) z P_{ba}^{R}\left(z, \alpha_{s}\left(b\left(z\right)^{2}\mu'^{2}\right)\right)\right)  . 
\end{eqnarray}
The initial evolution scale in Eq.~(\ref{eqn:PB})   is denoted by $ \mu_{0}$.

\begin{figure}
\centering
\includegraphics[width=10cm]{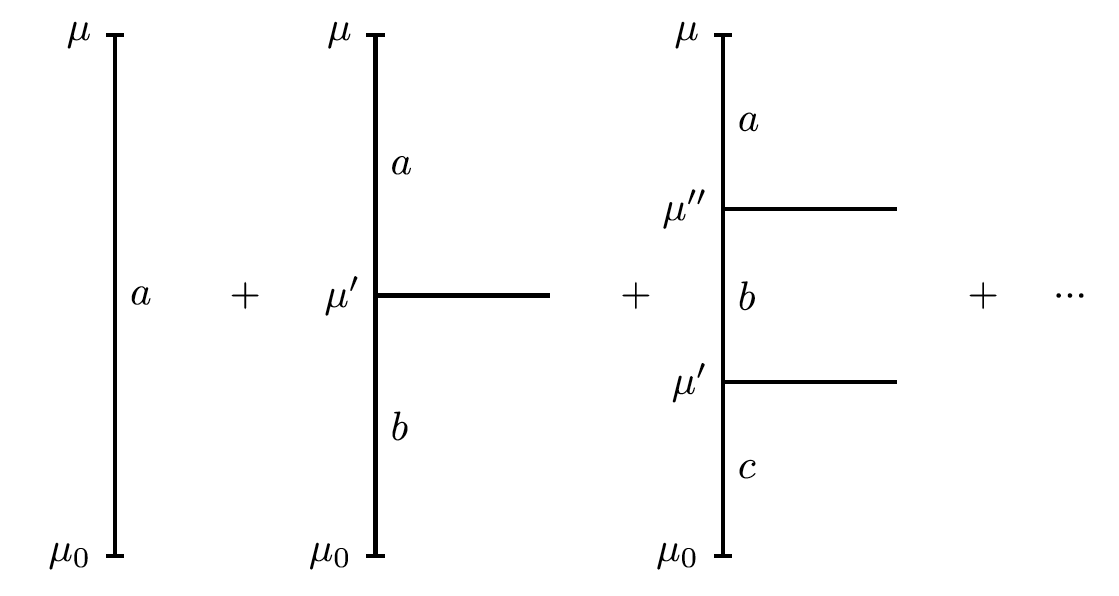}
\caption[Solution of the branching equation by iteration.]{Solution of the branching equation by iteration.}
\label{fig:Cascade.png}
\end{figure}

An iterative Monte Carlo solution of Eq.~(\ref{eqn:PB}) is obtained in \cite{Hautmann:1704},  and is represented pictorially in Fig.~\ref{fig:Cascade.png}. Collinear PDFs can be obtained from Eq.~(\ref{eqn:PB})  as well by integration over the transverse momentum $\it{k}$. The distribution of flavor $a$ at scale $\mu$ is written, as a function of $x$ and $k$, as a sum of terms involving no branching between $\mu_{0}$ and $\mu$, then one branching, then two branchings, and so
forth. The transverse momentum $\it{k}$, in particular, arises from this solution by combining the intrinsic
transverse momentum (in the first term on the right hand side of Eq.~(\ref{eqn:PB}))  
with the transverse momenta
emitted in all branching.

Due to colour coherence, soft gluons fulfill 
angular ordering (AO)  in the evolution cascade,  
  with the angle of the
emitted gluon with respect to the beam axis increasing at each branching. 
The PB method
incorporates the AO  \cite{Dynamical} through i) the  relation between the branching 
scale $\mu^\prime$ and the transverse momentum  $q$  
of the emitted parton, $|q|=(1-z)\mu^{\prime}$; ii) the scale in the running coupling 
$\alpha_{s}(q^{2})=\alpha_{s}((1-z)\mu^{\prime2})$;  iii) the resolution scale 
$z_m=1-{q_{0}}  / {\mu^{\prime}}$, where $q_0$ is the minimum transverse momentum 
with which a parton can be resolved.  
  Based on four-momentum conservation,  the constraint on the minimum transverse momentum of the emitted parton $q_{0}$ results, using the ordering relation, 
   in the maximum $\it{z}$ value, $z_{m}$, as a function  of the branching scale. Taking  $q_{0} $ to be large enough compared to $ \Lambda_{\mathrm{QCD}}$ allows one to stay in the weak coupling region avoiding the Landau pole of $\alpha_{s}$. Extending the work \cite{Bermudez},  in this study 
    we  present for the first time fits of  
   PB TMD parton densities with  
   $q_{0} $   larger than $ \Lambda_{\mathrm{QCD}}$ and dynamical 
   resolution scale.

\section{Fits at NLO with dynamical resolution scale}
\label{Fits at NLO with dynamical resolution scale}
\begin{figure}
\centering
\includegraphics[width=20pc]{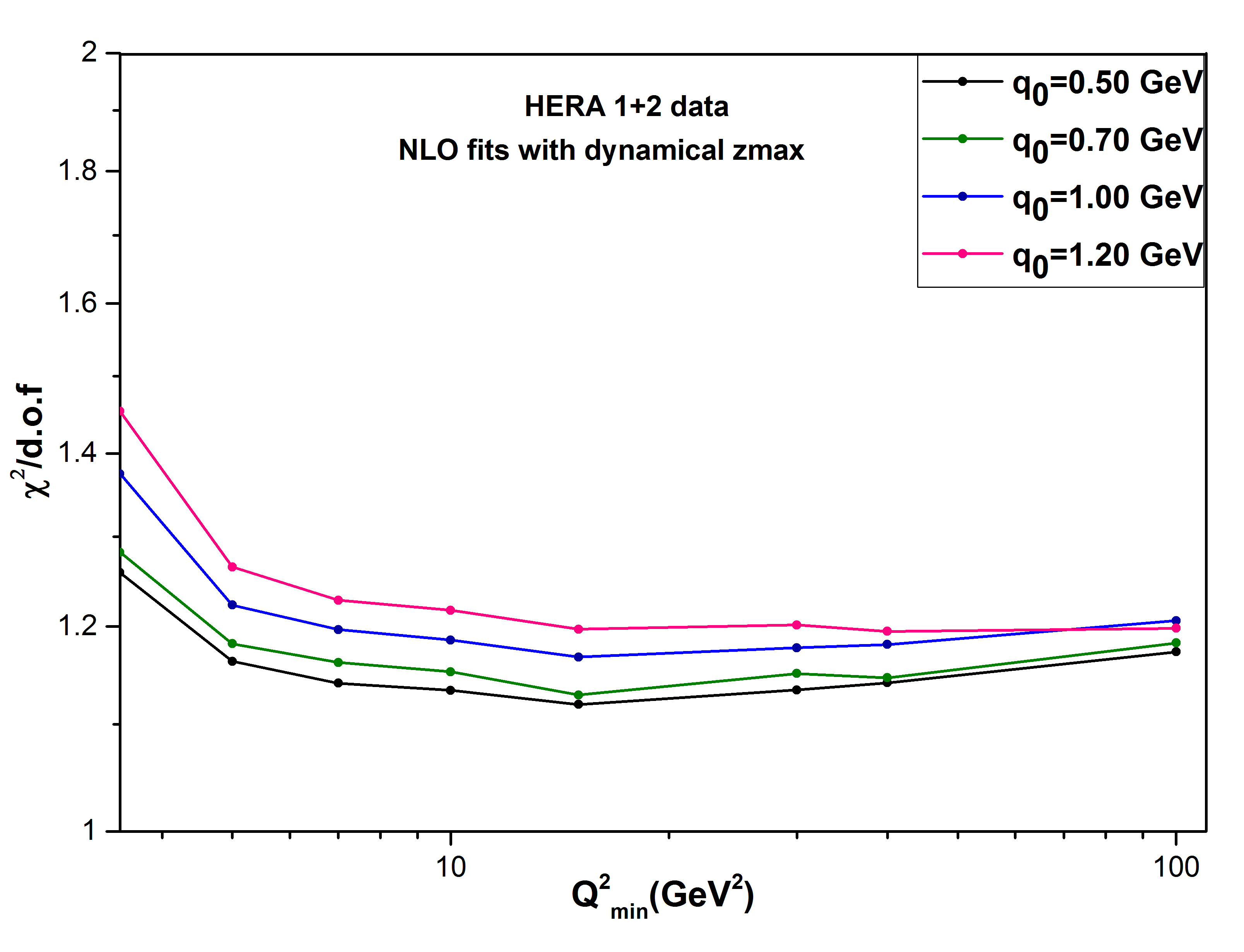}
\caption[The fit results with dynamical $z_m$ at NLO with HERA 1 + 2 data set, using xFitter]{The fit results with dynamical $z_m$ at NLO with HERA 1 + 2 data set, using xFitter}
\label{fig:plot1-1.png}
\end{figure}

In this section, numerical results for fits with dynamical resolution scale at NLO are discussed. The fits to inclusive DIS cross section combined H1 and ZEUS measurements \cite{Herapdf} are performed in a wide range of  $Q^{2}$ and $\it{x}$ using $\chi^{2}$ minimization, by means of the open-source QCD platform 
xFitter~\cite{HERAFitter,xFitterDevelopersTeam:2022koz}. 
We follow the same strategy  
as in  \cite{Bermudez}
for parameterization, systematic and experimental uncertainty calculations,  
use of NLO coefficient functions  and heavy flavour treatment. The experimental uncertainties on each of the fitted parameters are calculated within the xFitter package. 
The model uncertainty is obtained by varying ${m_{c}}$, ${m_{b}}$ and initial evolution scale. In Fig.~\ref{fig:plot1-1.png} results for the  ${\chi^{2}} / {d.o.f}$ as a function of  the minimum $Q^{2}$ of the data included in the fit are shown. In this figure, different fits with different ${q_{0}}$ values are plotted. It is observed that for each value of ${q_{0}}$ there is a reasonably good $\chi^{2}$ in a wide range of $Q^{2}$, with the  lowest 
${q_{0}}$ giving the best ${\chi^{2}} / {d.o.f}$.  

\begin{figure}
\centering
\includegraphics[width=35pc]{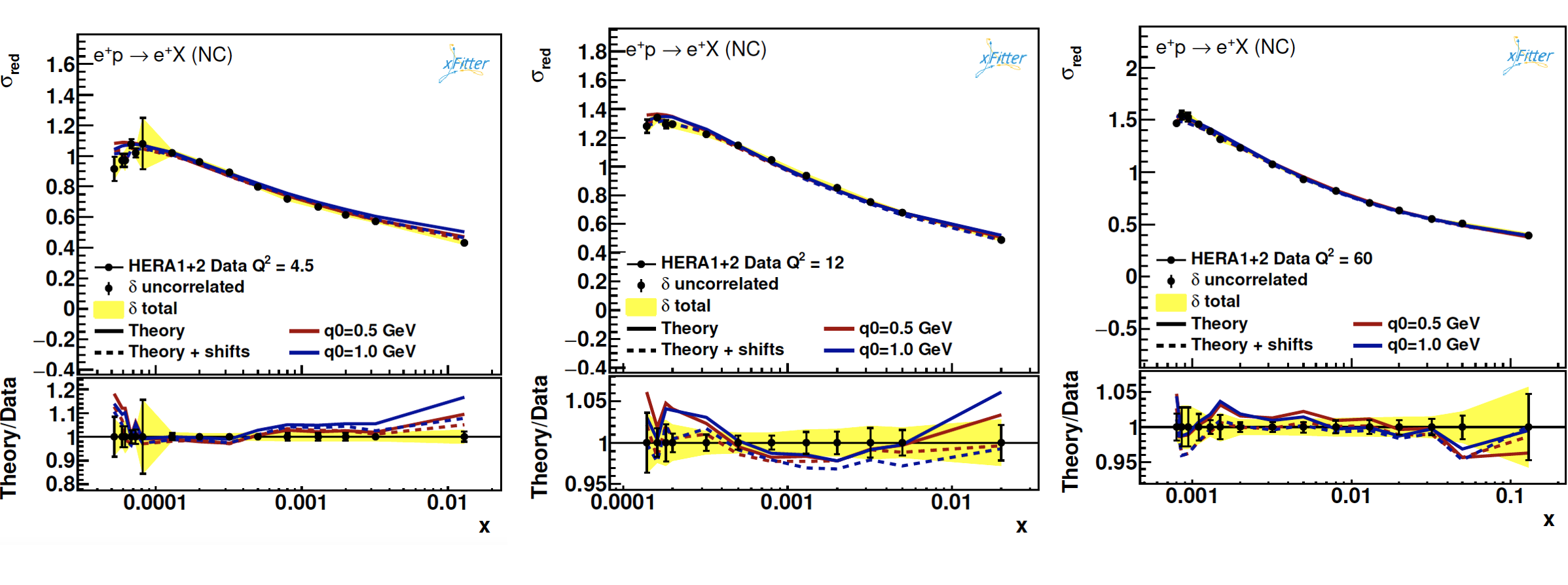}
\caption[Measurement of the reduced cross section obtained at HERA compared to predictions using ${q_{0}=0.5}$ and ${q_{0}=1.0}$.]{Measurement of the reduced cross section obtained at HERA compared to predictions using ${q_{0}=0.5 GeV}$ and ${q_{0}=1.0 GeV}$.}
\label{fig:reduced}
\end{figure}

\afterpage{\clearpage}
In Fig.~\ref{fig:reduced}, predictions for the inclusive DIS cross section from ${q_{0}=0.5 GeV}$ and ${q_{0}=1.0 GeV}$ are shown and compared with
the measurements from HERA \cite{Herapdf} for different values of the evolution scale $\mu^{2}=Q^{2}$. The agreement with data is excellent for these two different $q_{0}$s.

\begin{figure}
\centering
\includegraphics[width=30pc]{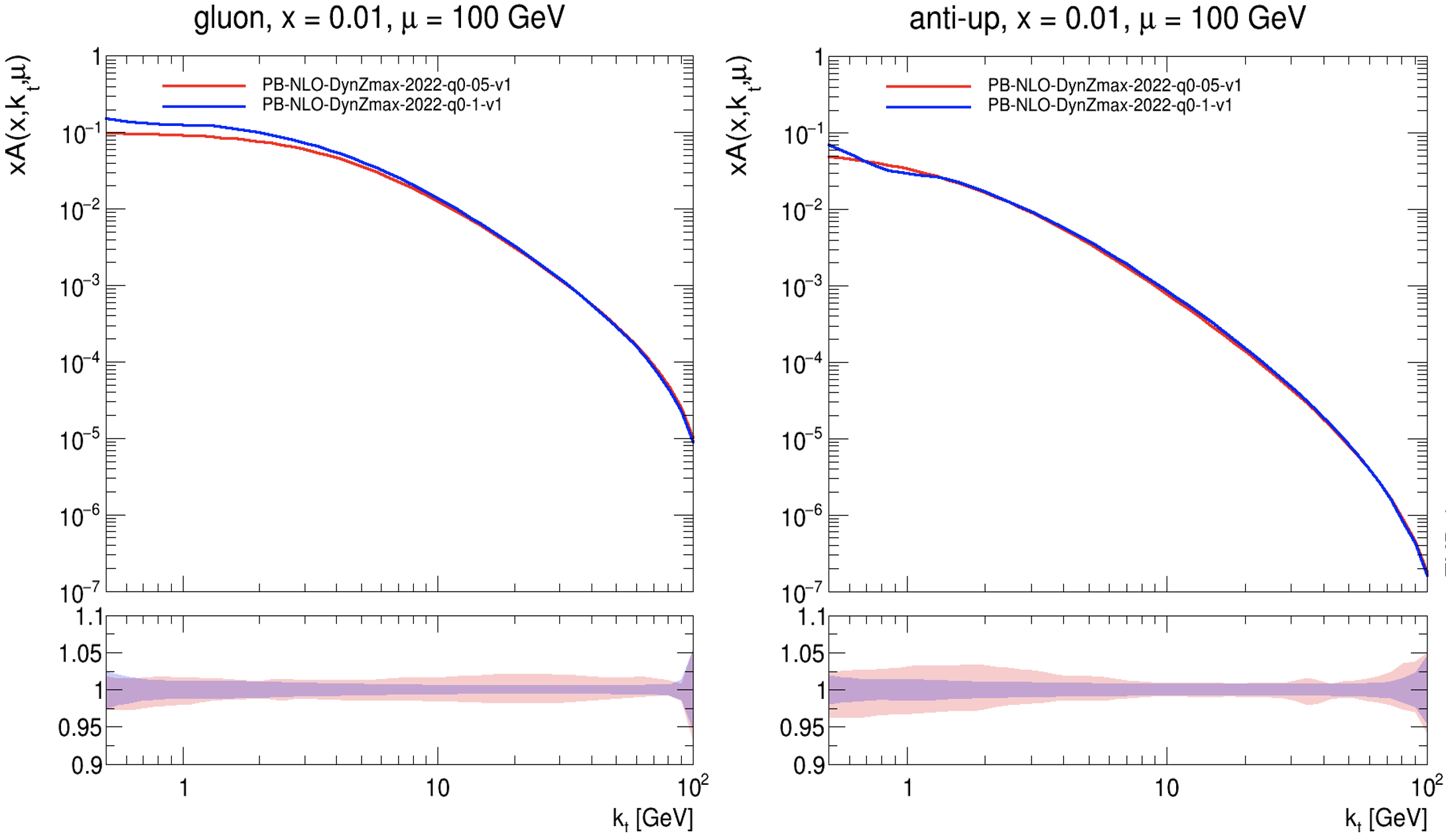}
\caption[Transverse Momentum Dependent parton densities for $\bar{u}$ and gluon with ${q_{0}=0.5 GeV}$ and ${q_{0}=1.0 GeV}$ as a function of $k_{t}$ for $\mu=100 \mathrm{GeV}$ at $x=0.01$. In the below panel, the relative uncertainty coming from the total of experimental and model uncertainties are shown.]{Transverse Momentum Dependent parton densities for $\bar{u}$ and gluon from ${q_{0}=0.5 GeV}$ and ${q_{0}=1.0 GeV}$ as a function of $k_{t}$ for $\mu=100 \mathrm{GeV}$ at $x=0.01$. In the below panel, the relative uncertainty coming from the total of experimental and model uncertainties are shown.}
\label{fig:uncert}
\end{figure}

In Fig.~\ref{fig:uncert} the gluon and $\bar{u}$ densities as a function of the transverse momentum are shown at $\mu=100 \mathrm{GeV}$ and $x=0.01$, for 
${q_{0}=0.5 GeV}$ and ${q_{0}=1.0 GeV}$, 
 together with the uncertainty band coming from the experimental and  model sources.

\section{Summary and outlook }
\label{Summary and outlook }

The PB method \cite{Hautmann:1708, Hautmann:1704}
has been used for angular-ordered TMD evolution, to obtain fits to the high precision
DIS structure function measurements, including for the
first time the effects of dynamical, i.e. branching-scale dependent, resolution scales. The fits have been  obtained at NLO accuracy in the strong coupling. Uncertainties on 
TMD distributions, including experimental and model sources, have been determined. 
Good agreement with the measurements is observed
when angular ordering is applied. 

The NLO TMD parton distributions determined from these fits, with the associated uncertainties, will be released this year, and will be made 
available in the TMDlib library~\cite{Abdulov:2021ivr,Hautmann:2014kza}. 

These distributions can be used for physics studies 
in hadronic collisions and phenomenological applications. 
It was noted in~\cite{BermudezMartinez:2019anj,BermudezMartinez:2020tys} 
that the low transverse-momentum part of the 
Drell-Yan lepton-pair production spectrum is sensitive to 
angular ordering effects in the QCD running coupling, and 
an analogous observation was made~\cite{Abdulhamid:2021xtt} 
for di-jet angular correlations at large azimuthal angles. The results presented in this work will enable investigations of 
angular ordering effects in soft-gluon resolution scales as well. 

Also Drell-Yan + jets final states can be studied with 
PB TMDs, e.g. by NLO matching~\cite{Yang:2022qgk} 
or multi-jet merging~\cite{Martinez:2021chk} approaches. For 
such studies, it will be possible to use the TMD distributions 
obtained in this work along with the Monte Carlo event 
generator~\cite{Baranov:2021uol} implementing the PB method. 
Dynamical resolution scale effects are currently being 
examined also in the context 
of generalized PB evolution equations involving TMD splitting 
functions~\cite{Hautmann:2022xuc,Keersmaekers:2021arn} 
defined by high-energy factorization~\cite{Catani:1994sq}.    

A future development of the work in this paper will be to 
extend the fits to larger data sets. This will apply both to 
data from current experiments such as LHC and fixed-target 
experiments, and to data from future collider programs 
such HL-LHC~\cite{Azzi:2019yne}, 
EIC~\cite{Proceedings:2020eah}, LHeC~\cite{LHeC:2020van}.

\section*{Acknowledgements}
The results presented in this article were obtained in collaboration with F. Hautmann, H. Jung, L. Keersmaekers, A. Lelek, S. Taheri Monfared. S.Sadeghi Barzani acknowledges funding by The University of Antwerp Research Fund (BOF).

\nolinenumbers


\begin{thebibliography}{99}




\bibitem{:Luisoni}
G. Luisoni, S. Marzani, J. Phys. G 42 (2015) 103101, arXiv:1505.04084.



\bibitem{:R. Angeles-Martinez}
R. Angeles-Martinez, et al., Acta Phys. Pol. B 46 (2015) 2501, arXiv:1507.05267.


\bibitem{Hautmann:1704}
F. Hautmann et al., Phys. Lett. B 772 (2017) 446, arXiv:1704.01757.

\bibitem{Hautmann:1708}
F. Hautmann et al., J. High Energy Phys. 01 (2018) 070, arXiv:1708.03279.



\bibitem{B.R. Webber}
B.R. Webber, Ann. Rev. Nucl. Part. Sci. 36 (1986) 253.


\bibitem{R.K. Ellis}
R.K. Ellis et al., QCD and Collider Physics, Cambridge University Press, 2003.     

    
\bibitem{A. Bassetto}
A. Bassetto, M. Ciafaloni, G. Marchesini, Phys. Rep. 100 (1983) 201.



\bibitem{Y.L. Dokshitzer}
Y.L. Dokshitzer, V.A. Khoze, S.I. Troian, A.H. Mueller, Rev. Mod. Phys. 60 (1988) 373.



\bibitem{G. Marchesini}
G. Marchesini, B.R. Webber, Nucl. Phys. B 310 (1988) 461.



\bibitem{S. Catani}
S. Catani, B.R. Webber, G. Marchesini, Nucl. Phys. B 349 (1991) 635.


\bibitem{F. Hautmann:hep -ph /0702196}
F. Hautmann, Phys. Lett. B 655 (2007) 26, arXiv:hep-ph/0702196.    


\bibitem{F. Hautmann:1407}
F. Hautmann, H. Jung, S.T. Monfared, Eur. Phys. J. C 74 (2014) 3082, arXiv:1407.5935.


\bibitem{Herapdf}
ZEUS, H1 Collaboration, H. Abramowicz et al., Eur. Phys. J. C75  (2015) 580,  
arXiv:1506.06042.

\bibitem{Dynamical}
F. Hautmann et al., Nucl. Phys. B949 (2019)  114795, arXiv:1908.08524.


\bibitem{Bermudez}
A. Bermudez Martinez et al., Phys. Rev. D99 (2019)  074008, arXiv:1804.11152.



\bibitem{HERAFitter}
S. Alekhin et al., Eur. Phys. J. C75 (2015)  304, arXiv:1410.4412.

\bibitem{xFitterDevelopersTeam:2022koz}
H.~Abdolmaleki \textit{et al.} [xFitter Developers' Team],
arXiv:2206.12465 [hep-ph].


\bibitem{Abdulov:2021ivr}
N.~A.~Abdulov \textit{et al.}, 
Eur. Phys. J. C \textbf{81} (2021)  752  
[arXiv:2103.09741 [hep-ph]].

\bibitem{Hautmann:2014kza}
F.~Hautmann \textit{et al.}, 
Eur. Phys. J. C \textbf{74} (2014) 3220  
[arXiv:1408.3015 [hep-ph]].

\bibitem{BermudezMartinez:2019anj}
A.~Bermudez Martinez \textit{et al.}, 
Phys. Rev. D \textbf{100} (2019) 074027
[arXiv:1906.00919 [hep-ph]].

\bibitem{BermudezMartinez:2020tys}
A.~Bermudez Martinez \textit{et al.}, 
Eur. Phys. J. C \textbf{80} (2020)  598
[arXiv:2001.06488 [hep-ph]].

\bibitem{Abdulhamid:2021xtt}
M.~I.~Abdulhamid \textit{et al.}, 
Eur. Phys. J. C \textbf{82} (2022)  36
[arXiv:2112.10465 [hep-ph]].

\bibitem{Yang:2022qgk}
H.~Yang   \textit{et al.}, 
arXiv:2204.01528 [hep-ph].

\bibitem{Martinez:2021chk}
A.~Bermudez Martinez, F.~Hautmann and M.~L.~Mangano,
Phys. Lett. B \textbf{822} (2021) 136700
[arXiv:2107.01224 [hep-ph]].

\bibitem{Baranov:2021uol}
S.~Baranov    \textit{et al.}, 
Eur. Phys. J. C \textbf{81} (2021)  425
[arXiv:2101.10221 [hep-ph]].

\bibitem{Hautmann:2022xuc}
F.~Hautmann, M.~Hentschinski, L.~Keersmaekers, A.~Kusina, K.~Kutak and A.~Lelek,
arXiv:2205.15873 [hep-ph].

\bibitem{Keersmaekers:2021arn}
L.~Keersmaekers,
arXiv:2109.07326 [hep-ph].

\bibitem{Catani:1994sq}
S.~Catani and F.~Hautmann,
Nucl. Phys. B \textbf{427} (1994) 475 
[arXiv:hep-ph/9405388].

\bibitem{Azzi:2019yne}
P.~Azzi \textit{et al.}, 
CERN Yellow Rep. Monogr. \textbf{7} (2019)  1 
[arXiv:1902.04070 [hep-ph]].

\bibitem{Proceedings:2020eah}
Y.~Hatta  \textit{et al.}, 
arXiv:2002.12333 [hep-ph].

\bibitem{LHeC:2020van}
P.~Agostini \textit{et al.} [LHeC and FCC-he Study Group],
J. Phys. G \textbf{48} (2021)  110501
[arXiv:2007.14491 [hep-ex]].


\end{thebibliography}
\end{document}